Nashid Shahriar[1], Mahfuza Sharmin[1], Reaz Ahmed[1], and Raouf Boutaba[2]

[1]Department of Computer Science and Engineering
Bangladesh University of Engineering and Technology, Dhaka, Bangladesh
{nshahriar, sharmin, reaz}@cse.buet.ac.bd
[2]School of Computer Science, University of Waterloo Ontario, Canada
rboutaba@bbcr.uwaterloo.ca

**Abstract.** This paper discusses an efficient approach to design and implement a highly available peer-to-peer system irrespective of peer timing and churn. Although peers in P2P system join or leave at whim, it has been found that most of the peers follow diurnal pattern of availability governed by the time of day effect. When considering a global P2P system, the cyclic behavior of peers situated on different time zones can be found complementary of one another. Our approach utilizes the diurnal pattern of globally dispersed peers to develop a grouping strategy. The objective of each group is to ensure 24x7 data availability within the group. To represent availability pattern we propose to divide 24 hours of a day into multiple slots and then to express the availability of a particular peer in each slot. In our approach, each peer collects slot availability information of a number of peers and forms small groups of 4 to 8 peers in such a way that the combined availability in each slot within a group is close to 100%. Simulation results show that our protocol converges fast and ensures high availability for each group with minimal overhead.

## 1 Introduction

There has been a rapid increase in the popularity of the Peer-to-Peer(P2P) systems during the last decade. Such systems typically lack dedicated infrastructure, centralized administration, prerequisite conditions, but rather depend on the voluntary participation of individual computers often referred to as peers. These properties of P2P systems attract enormous number and heterogenous types of peers distributed across the world to participate and contribute their otherwise unused resources. The utilization of unused resources and collocation of computation power of end computers have become the main strength of P2P systems which offer low cost, large capacity, and enormous computation power. As the geographically dispersed computers may be separately owned and managed, peers can join or leave the system autonomously and may be available for arbitrary period of time. These dynamics of peer participation also known as churn constitute an inherent characteristic of P2P systems[15]. The permanent flow of peer connections and disconnections and arbitrary duration of presence may severely hamper the data availability in a P2P system. It is still a challenging issue to design and implement a highly available P2P system without incurring significant replication overhead.

Availability is the property which ensures that data can be retrieved at any moment, i.e., being always available. It can be expressed by the probability that the data can be retrieved at a given time [3]. The basic strategy for developing a highly available P2P system is replication, i.e., to keep redundant data on multiple peers. A P2P system utilizing data replication may tolerate failure of a few peers as the alive peers containing redundant data may provide the required data. A number of approaches for data replication in P2P systems can be found in the literature. These approaches vary in the type of redundancy, method of data regeneration, and the timing and number of peers for storing redundant data. In terms of method, redundancy can be achieved either by replication of the original data or encoding and fragmenting of encoded data such that not all fragments are needed to reproduce data [2]. Replication of data is mainly done in two ways: reactive [4] or proactive [6], [5]. In both approaches, availability is increased by increasing the number of replicas.

However, increasing the number of replicas comes at the cost of a higher data replication traffic, higher storage requirement, and increased cost and complexity of update propagation.

Achieving high data availability with minimal number of replicas is an essential design requirement in a P2P system. Existing approaches for replication utilize information like peers' previous availability pattern [17], lifespan distribution [24], machine availability, [9], [10], Mean Time to Failure [3], up time score [23], recent up time [19], [7], application specific availability [8], session time and churn [13], probabilistic model [4], [11]. However these approaches require a high number of replicas to attain moderate data availability. They are biased towards highly available peers, which skews the distribution of free space among machines and creates congestion towards these peers thereby limiting the availability. Another problem lies in gathering availability information from the large number of peers in unstructured P2P systems with no centralized component. In this paper, we propose a protocol that collects peers' availability information in a fast and efficient way and ensures high availability with a small number of replicas while avoiding any bias.

Our approach for ensuring higher availability with lower replication utilizes the availability patterns of geographically distributed peers. Indeed, previous studies [16], [17], [12], [18], [10] show that availability behavior in most of the peers in a P2P system oscillates over a 24-hour period governed by the time of day effect. This indicate that the peers follow a diurnal pattern [14]. When considering a global P2P system, such cyclic behavior of peers situated in different time zones can be found very diverse. Our approach utilizes the phase relationships of diurnal patterns of globally dispersed peers to achieve the aforementioned goal of ensuring high availability with minimal replication.

## 2 Motivation

Measurement studies of popular P2P systems indicate a great level of heterogeneity among the individual peers in terms of availability. In spite of such heterogeneity, studies on P2P systems have been able to find the inherent availability characteristics of the participating peers. Based on their study of the Gnutella and Napster networks, Douceur et. al. [12] concludes that the computers at network edge (i.e., peers) exhibit recurring and cyclic internet connectivity pattern. This pattern is mostly governed by individual preference, work schedule, time of day etc. Some peers may show high availability during peak hours which may be the first quarter of the day and low availability in all other time, whereas some other peers may show low availability in most of the time except the middle few hours of the peer's local time. Interestingly, such diurnal behavior governed by the time of the day cycles through a long period of time unless there is some major change in a peer's location or habit. The time-of-day effect in peer availability has been observed in the analysis of the Gnutella network conducted in [14] and [16]. Bhagwan et al. [1] observed the diurnal pattern in peer availability during their study of the Overnet file-sharing network. Le Blond et al. [17] observed the global diurnal patterns of around 100,000 peers is clearly visible in their analysis of the traces from the eDonkey network.

Peers in a P2P system are usually distributed across the world in different time zones. When considering a global P2P system under a universal time standard i.e. UTC, the cyclic behavior of the peers situated on different time zones can be found to be totally or partially overlapping or total or partial complementary. For example, two cyclic peers that are usually down during the night, but located in Dallas and Dhaka, show complementary availability pattern due to the 12-hour clock difference. Even the peers located in the same time zone may show diverse availability pattern depending on the individual's internet habit or job function. If we consider all the peers across the world, their peak availability will be distributed across different time periods of the day in the UTC standard and for peers in the nearby time zones there will be some overlapping portions. This fact is illustrated in Figure 1, where peers A, B, C, and D having similar daily Internet habit and located in four different time zones can togather provide improved availability around the clock. In our proposed scheme we take advantage of this kind of behavior to improve availability with smaller replication overhead.

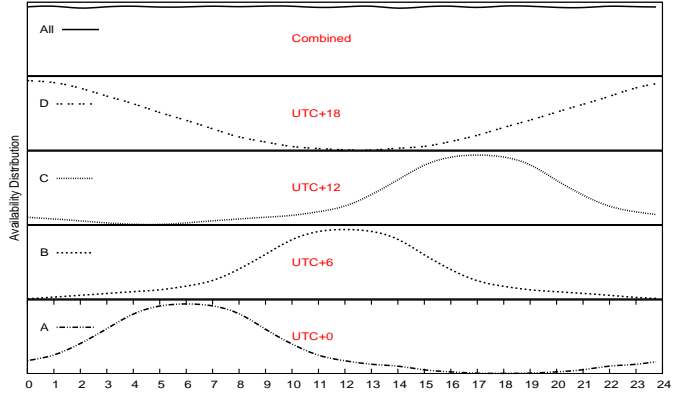

**Fig. 1.** Global Availability Pattern

## 3 Proposed Strategy

In our proposed strategy, peers having complementary availability behaviors form a group. Low availability of a peer in some time slot is compensated by another member of the group with high availability in that time slot. Such a group ensures that at least one member of the group is available with high probability at any given time. Collectively, it ensures overall increase in the system's availability. Now, the challenge is how to form such groups in a distributed manner for an unstructured P2P system. If we were able to know the availability pattern of all the peers, it would be mere easy to form such groups optimizing complementary patterns. But, collecting and searching availability patterns of all the peers is not feasible without any centralized mechanism. So, we propose a gossip based approach to construct a self-organizing gradient topology [19] of optimistic availability patterns using our defined metrics. The information contained in the gradient topology is used to form the desired groups.

At the beginning, only individual peers are present in the system but later both isolated peers and groups may be present in the system. For consistency, our protocol treats individual peers and multi-member groups identically. A single peer constitutes an isolated group where the peer is the sole member. We call it Singleton Group and rest of the groups as multi-group. When a peer joins the system it will get a *GroupID* from the system and when two groups with separate *GroupID* merges they will be identified by a new *GroupID*. Each group will have a representative peer, which is the only peer in the group if it is a singleton or the currently alive peer if it is a multi-group. For the case of multi-group, our approach ensures that at least one peer is available across different periods along the time. During a period, when more than one members are concurrently alive, the peer that joined last will be chosen as the representative or we can use any other leader selection mechanism.

After the formation of a group, any information of any member of the group will be replicated to all other members of the group. As our protocol try to minimize number of group members, such flooding would not cost much. The group as a whole is responsible for ensuring the availability of the document around the clock. The query mechanism for the document placement need not to be changed a lot other than considering the group instead of the peer as the basic unit. To do so, our protocol needs to map the document ID with its underlying group ID instead of a bunch of peer IDs. A similar routing mechanism where small groups are considered as the basic unit of the Chord [26] model is described in [27]. The peer looking for the document will communicate using the group ID and a representative will provide the document on behalf of the group. This way the proposed protocol, ensures with a high probability that a representative peer will be available.

### 3.1 Availability vector

Our proposal takes into account previous history of availability of peers, which globally obey some patterns. To represent that pattern, we divide the 24 hours of a day into multiple, say $K$, slots of equal length $l$. Accordingly, we have a total of $K = 24/l$ slots for each peer or group. We propose to represent availability $a_{ik}$ of a peer $P_i$ in a particular slot $k$, by the probability of $P_i$ to be available at $k_{th}$ time slot of the day based on its historical behavior. By gathering the peer's most common availability information throughout all the slots, we get $a_{ik}$ for all the $K$ slots of a day. We then represent availability behavior of a peer as a $K$ dimensional vector, named Availability Vector $(\mathring{A})_i$, by the following equation:

$$(\mathring{A}_i) = \{a_{i1}, a_{i2}, ..., a_{ik}, ..., a_{iK}\} \tag{1}$$

Peer $P_i$ can easily compute its Availability Vector $(\mathring{A})_i$ by recording its Internet connectivity history for a sufficient period of time. Once computed, the $(\mathring{A})_i$ can be updated through periodic observation on the availability of the peer. Accurate calculation of $(\mathring{A})$ considering sudden failures of peers in the middle of slots, uptime distribution, mean session time is beyond the scope of this paper. Here, we assume that the P2P software running on each peer deploys a standard method to compute the peer's availability vector simply from its previous history.

We, now define the $(\mathring{A})$ for a multi-group. The equation is the same but the meaning of individual components is now different as more than one peer are involved. For multi-group, each $a_{pk}$ of $(\mathring{A})_p$, represents the probability of at least one of the members of the group $G_p$ to be available during the $k_{th}$ slot of the day. When a peer joins a group, its individual $(\mathring{A})$ becomes invalid and it will capture the $(\mathring{A})$ of the group as its own.

### 3.2 Metric for Group Formation

We define, the contribution $C_{i,j}$ between two groups $G_i$ and $G_j$ as the improvement on availability after merging $G_i$ and $G_j$ in a new group. This contribution is the most vital metric in our protocol because when forming groups, a group will use this metric to build the gradient topology and to select its group mate. We develop two equations to calculate contribution from two $(\mathring{A})$s of the two participating groups. Before posing the equations, we define some terminology that will be used later to explain the equations.

Let, $C_{ijk}$ denotes the slot wise contribution of any two groups $G_i$ and $G_j$ i.e. the contribution of $G_i$ and $G_j$ only at slot $k$. The size of the group $G_i$ is symbolized by $|g_i|$. $|g_i \cup g_j|$ is the size of the new group consisting of the former two groups. The joint availability of two groups $G_i$ and $G_j$ at slot $k$ is denoted by $J_{ijk}$. We define the joint availability $J_{ijk} = a_{ik} * a_{jk}$. Now, we find the value of $C_{i,j}$ as follows:

$$C_{i,j} = \sum_{k=1 to K} \frac{C_{ijk}}{|g_i \cup g_j|} \quad (2)$$

where,

$$C_{ijk} = \begin{cases} J_{ijk}^{(\frac{a_{ik}}{a_{jk}})} - J_{ijk} \text{ if } a_{ik} \leq a_{jk}; \\ J_{ijk}^{(\frac{a_{jk}}{a_{ik}})} - J_{ijk} \text{ if } a_{ik} > a_{jk}. \end{cases}$$

We now explain the motivation behind proposing such equation. It is typical that, a group having larger number of members will provide higher availability than a group having a single member or smaller number of members. During candidate selection, a peer will find the larger group more attractive than a smaller one. This may lead to a situation where some groups have large number of members while keeping some others just as *Singleton* groups. Hence and besides ensuring higher availability our strategy also aims at keeping the group size as small as possible. Therefore to minimize the selection of larger groups, the term $\frac{1}{|g_i \cup g_j|}$ has been introduced in 2 as a factor while summing up the slot wise contribution.

As discussed in previous subsection, for any two groups, $G_i$ and $G_j$, more complementary slots should contribute more in the equation and to reflect this the value of $J_k$ is deducted from the term $J_{ijk}^{(\frac{a_{ik}}{a_{jk}})}$ (or $J_{ijk}^{(\frac{a_{jk}}{a_{ik}})}$). Here, larger difference between $a_{ik}$ and $a_{jk}$ gives smaller ratio as exponent which in turn powers the term, $J_{ijk}^{(\frac{a_{ik}}{a_{jk}})}$ (or $J_{ijk}^{(\frac{a_{jk}}{a_{ik}})}$) to larger value and results in larger slot contribution $C_{ijk}$. To summarize, consider three groups $G_i$, $G_j$ and $G_l$ having availability $a_{ik}$, $a_{jk}$ and $a_{lk}$ at slot $k$. For peer $G_i$, $G_j$ is more attractive than $G_l$ at slot $k$ if $a_{jk} > a_{lk}$.

As the exponent of joint probability, $J_{ijk}$, $|a_{ik} - a_{jk}|$ can be used instead of the ratio $\frac{a_{ik}}{a_{jk}}$ (or $\frac{a_{jk}}{a_{ik}}$) to produce the same effect. But the ratio has been chosen to distinguish the case where two pairs of values (single element of availability vector) have same difference but minimum (or maximum) of the pairs are different. In this cases pairs having smaller minimum (or maximum) should contribute more in the equation. Suppose for groups $G_i$ and $G_j$, $|a_{ik} - a_{jk}| = |a_{il} - a_{jl}|$ and minimum$(a_{ik},a_{jk})$ < minimum$(a_{il},a_{jl})$ (or maximum$(a_{ik},a_{jk})$ < maximum$(a_{il},a_{jl})$), where $a_{ik}$, $a_{jk}$, $a_{il}$ and $a_{jl}$ are slot availability at slot $k$ and $l$. Now, the minimum$(a_{ik},a_{jk})$ of slot $k$ demands prior compensation. To satisfy this requirement we have oriented in such a way so that slot $k$ contributes more than slot $l$ in the final equation of $C_{i,j}$.

The explanation in favor of calculating contribution by another equation is relatively straightforward. If the two groups $G_i$ and $G_j$ are merged into a larger group $G_p$, the former group $G_i$ (or $G_j$) should be benefitted by the amount $U_{ip}$ ($U_{jp}$). Mathematically,

$$U_{ip}(U_{jp}) = \sum_{k=1 to K} (\alpha_{pk} - \alpha_{ik}(\alpha_{jk}))$$

Here, $\alpha_{pk}$ is the $\alpha$-availability of new group to be formed at slot $k$ and $\alpha_{ik}$ (or $\alpha_{jk}$) is the $\alpha$-availability of one of the former groups at slot $k$. We define the $\alpha$-availability of a group as the availability of at least $\alpha$ member(s) of that group at that slot. To ensure at least one member of a groups to be available across the time we need to optimize $\alpha = 1$ that is 1-availability. Hence the following equation suffices to find the contribution $C_{i,j}$ satisfactorily. The argument in favor of using the the term $\frac{1}{|g_i \cup g_j|}$ as factor is the same as discussed earlier.

$$C_{i,j} = \frac{(U_{ip} + U_{jp})}{|g_i \cup g_j|} \quad (3)$$

### 3.3 Availability information

Peers in unstructured P2P systems are independent, have limited resources, keep little knowledge about the system and interact with a limited number of neighbors. In such a system, to find the current best candidate to group with, our protocol needs to devise a search strategy that is efficient, scalable and that converges fast without any centralized component. To keep the search space relatively smaller, we propose to maintain a local list of current best candidates named as *knownlist* in each peer in a gradient manner. The search algorithm exploits the above list to achieve a significantly better search performance than traditional search techniques, such as random walking, which require the communication with potentially all peers in the system. The absence of centralized component requires that the construction and maintenance of the local list should be self-organized. In our protocol, we propose a gossip based information exchanging method named exploration to generate a list of current best candidates under a completely decentralized environment. Each peer after joining the system for the first time or after its offline cycle, starts its journey through exploration. The peer as a representative of some group periodically runs exploration during the time it remains present in the system.

During exploration, the representative peer of a group exchanges with its neighboring groups and group mates their availability information. We define two groups $G_i$ and $G_j$ as neighbors if any member of $G_i$ keeps logical connection with any member of $G_j$. The *knownlist* of a group consists of selected GroupIDs whose availability information it has found to be advantageous in terms of contribution and their corresponding GroupSizes and ($\mathring{A}$)s. The groups in the *knownlist* can be either singleton or multi-group. Although a representative peer can gather availability information of a number of groups, it only keeps the predefined *knowncount* number of groups in its *knownlist* whose contributions are better in the current context. Apart from *knownlist*, each representative peer also stores ($\mathring{A}$) of the group in which it belongs, a list of neighboring groups with whom its underlying group has logical connection.

In the exploration phase, two neighboring groups $G_i$ and $G_j$ respectively gossip with each other to exchange their ($\mathring{A}$), grouping information and *knownlist*. To do so, the representative peer $P_i$ of exploring group $G_i$, sends *request* message to $G_j$ asking for $G_j$'s relevant information piggybacking its own bundle of information in the *request*. The representative peer $P_j$ on behalf of the receiving group $G_j$, sends its bundle of information in *reply*. At this stage, both $P_i$ and $P_j$ have received the ($\mathring{A}$)s of each others' *knownlist*. Both peers locally compute the contribution for each of the collected groups' ($\mathring{A}$)s using the previously described equations. Both peers then compare the contribution of each entry present in its own *knownlist*, with the newly computed contributions in its end and update its *knownlist* with the best set of groups having the highest contributions. The exploring peer then repeats the whole process with all of its neighboring groups. If the exploring peer is a member of multi-group, it also sends *request* message to its group mates in addition to its direct neighbors. By the end of its exploration, the group is expected to gather the information of best matching groups in its one or two hop neighborhood distance. In a similar way, all the groups currently present in the system perform exploration, update their *knownlist* with the best matching groups. As the exploration continues, discovery of best candidates expands from two hop to larger distances.

In our work we assume that each peer computes its $(\mathring{A})$ simply from its previous history and advertises it honestly when asked by any other peer. Relying on a peer itself to provide its availability information may create problems in the presence of malicious peers. These peers may deliberately misreport information if there is an incentive to do so. Tackling the untrusted behavior of peers is another research issue and can be investigated in future work.

## 3.4 Group Construction

A newly joined peer after gathering availability information and updating *knownlist* through exploration executes a grouping phase. The purpose of grouping is to form a group of peers so that the group ensures high availability throughout all the slots of a day utilizing the variation in the availability patterns of its members. Such a group is constructed incrementally i.e. forming groups with two single peers initially then growing in size up to the maximum allowable group size. In the later grouping phases, two non-singleton groups merge into a larger group such that resultant availability of the new group in all the slots increases from the availability of the former two groups by a sufficient margin.

In the grouping phase, each group $G_i$ executes $MakeGroup$ as shown in Algorithm 1 where its representative peer $P_i$ searches the entries in its *knownlist* to find its best matching group and invites it to merge with. To do so, $P_i$ picks the group $G_{max}$ with the highest contribution in its *knownlist* and sends *groupinvitation* message to $G_{max}$. Upon receiving *groupinvitation*, the representative peer $P_m$ respond on behalf of the invited group $G_m$. In our grouping policy, we do not allow a peer to be the member of more than one group at the same time. So, $P_m$ sends a *denial* message back to $G_i$ if $G_m$ is already involved in the formation of groups. Otherwise, $P_m$ compares the contribution of $G_i$ with the contributions of the entries in its *knownlist*. If the contribution of $G_i$ is still greater than those of the *knownlist* only then $P_m$ sends an *acceptance* message to $G_i$ otherwise it sends a *denial*. Upon receiving a *denial* from $G_m$, $P_i$ repeats the process by picking one of the rest of the groups in the *knownlist* in order of their contribution and inviting it as previously described. If none of them agrees to form a group, $G_i$ remains unchanged. On the other hand, getting the *acceptance* from $P_m$, $P_i$ initiates the activities of coalescing the two groups $G_i$ and $G_{max}$ into a larger group and the $MakeGroup$ for $G_i$ terminates.

At this stage both groups reach an agreement to form a new group and exchange some more information like $(\mathring{A})$, *knownlist*. The process of merging two groups into a new group is shown on Algorithm 3. The newly formed group gets a new $GroupID$ from the system. Then, $P_i$ calculates the new availability vector for the group, $(\mathring{A})_p$ using both of the $(\mathring{A})_i$ and $(\mathring{A})_j$. This new $(\mathring{A})_p$ is propagated to all members of the group and each member considers $(\mathring{A})_p$ as their individual $(\mathring{A})$ from now on.

## 3.5 Group Maintenance

As we discuss in the last section, any peer that joins the system for the first time, executes exploration and grouping in sequence to become a part of some group. After joining a group, a peer may execute grouping periodically as the representative of that group. While doing so, it only focuses on growing the group to attain a higher availability and not to leave the group. After some period, the peer may become offline according to its diurnal pattern and our grouping strategy ensures some other members of the group will be available on behalf of the group. After its offline period when the peer rejoins the system, it retains the membership of the same group as it was previously. The membership of a peer in a group can be changed only when there is some modification in the availability pattern of the peer. In such a case, a peer leaves the current group and the $(\mathring{A})$ of the group is modified accordingly. The leaving peer then executes exploration and grouping in sequence with its new $(\mathring{A})$ as if it joins the system for the first time.

**Algorithm 1** MAKEGROUP($g_i$)
─────────────────────────────────────────
1: If $|g_i| = 1$
2:     $p_i \leftarrow g_i.peer$
3: else
4:     $p_i \leftarrow$ a representative peer from $g_i$
5: $K_i = \{g_1, g_2, ....g_j\}$ where $g_j$ are the peers in the *knownlist* of $P_i$
6: For each $g_j$ in $K_i$
7:     compute $C_{i,j}$
8: Sort $K_i$ in order of decreasing $C_{i,j}$
9: while $K_i$ not empty
10:     $g_{max} \leftarrow g_j$, such that $C_{i,j}$ in $K_i$ is maximum
11:     Send *groupinvitation* to $g_{max}$
12:     Wait for $WAIT\_INTERVAL$
13:     if *acceptance* received
14:         $new\_group \leftarrow g_{max}$
15:         exit from loop
16:     else
17:         $K_i \leftarrow K_i - \{g_{max}\}$
18: Repeat
19: if *new_group* is not null
20:     $mergegroup(g_i, new\_group)$
21: else
22:     return

**Algorithm 2** REPLYINVITATION($g_i$, $g_m$, $C_{i,m}$)
─────────────────────────────────────────
1: If $|g_m| = 1$
2:     $p_m \leftarrow g_m.peer$
3: else
4:     $p_m \leftarrow$ a representative peer from $g_m$
5: if $g_m$ involved in grouping
6:     Send *denial* to $g_i$
7: else
8:     $C_{max} \leftarrow$ maximum value of contribution in the *knonwlist* of $p_m$
9:     $|g_i \cup g_m| \leftarrow |g_i| + |g_m|$
10:     if $C_{i,m} > C_{max}$ and $|g_i \cup g_m| <$ MAXGROUPSIZE
11:         Send *acceptance* to $g_i$
12:     else
13:         Send *denial* to $g_i$

**Algorithm 3** MERGEGROUP($g_i$, $g_m$)
─────────────────────────────────────────
1: $g_p \leftarrow GroupID$ of the new group assigned by the system
2: $(\mathring{A})_i \leftarrow$ availability vector of $g_i$
3: $(\mathring{A})_m \leftarrow$ availability vector of $g_m$
4: For each slot $k$ from 1 to $K$
5:     $(\mathring{A})_p[k] \leftarrow 1 - ((1 - (\mathring{A})_i[k])(1 - (\mathring{A})_m[k]))$
6: $memberlist(p) \leftarrow memberlist(g_i) \cup memberlist(g_m)$
7: $|g_p| \leftarrow |g_i| + |g_m|$

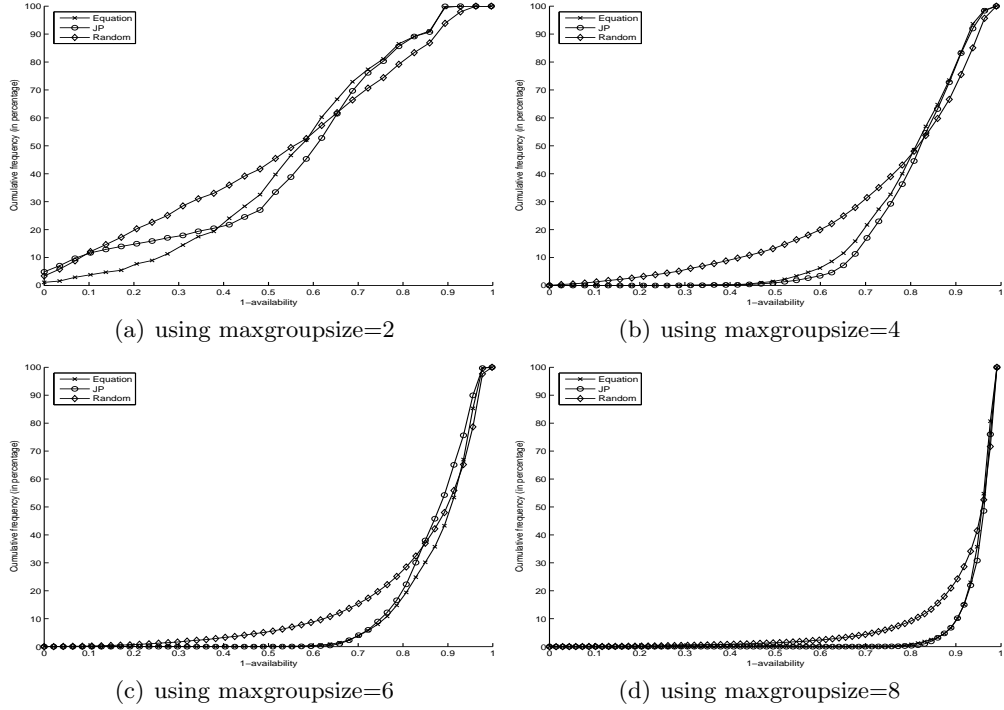

**Fig. 2.** Comparison with 1-availability

## 4 Experimental Results

To evaluate the performance of our protocol, we build a detailed simulation model of unstructured P2P network using C++. Using this model, we have implemented our strategy and conducted simulations with 10,000 peers as follows. In our environment, a peer has neighbor degree between 5 to 10. So, a peer has knowledge about only small portion of the network which is logical. To represent availability patterns of peers in a day we use 12 slots each of which is 2 hours long. The availability vector, $(\mathring{A})$, for each peer has been generated randomly considering the fact that each peer has a peak time in the 24 hours during which its availability will be high. These peak availabilities for different peers can be at any slot in the 24 hour duration as shown on Figure 1. The individual elements in $(\mathring{A})$ are either positively or negatively correlated for consecutive slots. During simulation peers keep at most 10 entries in their *knownlist*.

To illustrate the effectiveness of our proposed protocol, we examine the 1-availability of the groups constructed using both of our selection criteria. We also compare 1-availability of the groups constructed using our protocol with the groups formed in a random manner. According to the definition of $\alpha$-availability, 1-availability of a group at a slot is the probability of at least one member of the group to be available at that slot. Higher values of 1-availability for a group across all the slots are desired to assure a highly available system. The goal of our protocol is to make 1-availability of each slot of every group as high as possible. To verify whether the goal is fulfilled, all individual elements of the $(\mathring{A})$s of each group which are in fact 1-availability at different slots of every group need to be examined. To show the results, we have taken 1-availability values along x-axis and cumulative frequency (taken in percentage) of occurring those values in the slots of all groups after convergence along y-axis. Each element of $(\mathring{A})$ of each group actually falls in one of the $n$ buckets, where each bucket length is $\frac{1}{n}$ having the range of $jth$ bucket from $\frac{(j-1)}{n}$ to $\frac{j}{n}$. The value of $n$ is chosen according to Scott's rule [28]. The experiment is done for different maxgroupsizes. As shown in Figures 2, when groups are formed randomly larger percentage of slots provide smaller 1-availability values even though 1-availability of some of the groups reach to higher values. Figure 2 also shows that groups constructed with our protocol using both devised equations have very small percentage of slots having 1-availability of less than 0.6 and majority of the slots of the groups have 1-availability close to 1.

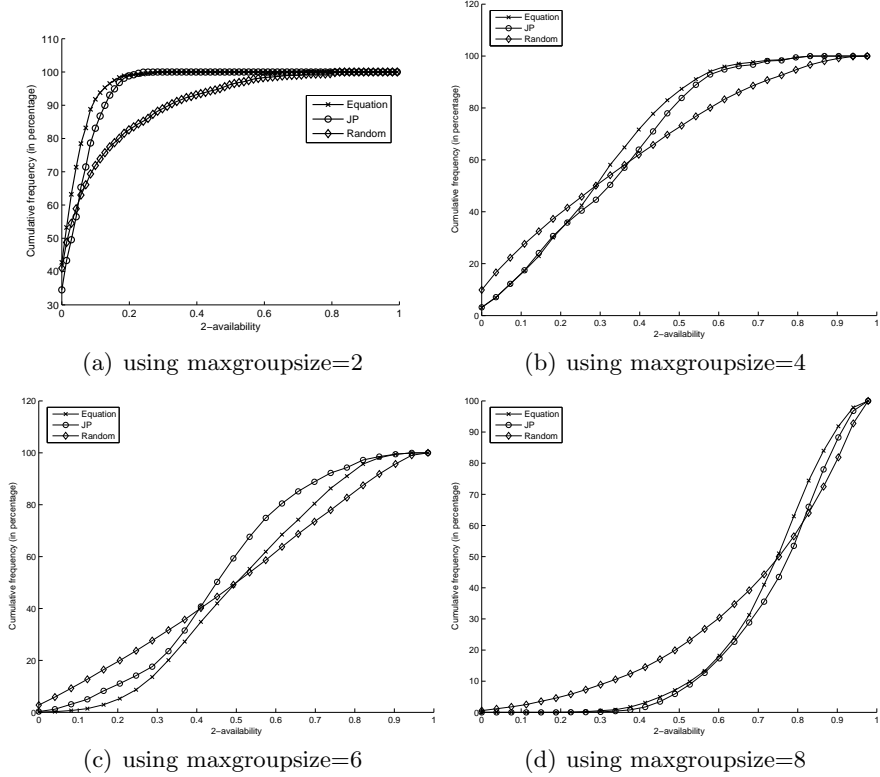

**Fig. 3.** Comparison with 2-availability

It is also visible from the figures that with the increase of the maximum allowable group size the distinction between group selection process decreases. This stems from the fact that having more members in a group increases the probability of finding at least one of them available at any time. This is the technique used by many traditional systems to improve availability but conflicts with our second goal of keeping the member count of a group as small as possible. Though Figure 2(d) is the clear winner, it warns us about this goal. Figure 2(c) with maximum group size of 6 is the best result that achieves both goals: improved availability at the cost of minimal redundancy. Although ensuring 1-availability is our prime concern, we find that groups formed using our protocol provide much higher 2-availability across the slots than the ones when groups are formed randomly thereby providing better reliability (see Figure 3).

## 5 Related Works

A number of approaches to improve availability can be found in the literature. To the best of our knowledge, very few approaches are proposed that try to optimize performance using the availability patterns of globally dispersed peers. Schwarz et all [23] propose to improve the availability of erasure coding scheme by utilizing cyclic behaviors of peers distributed across the world. They propose hill-climbing strategy to determine redundancy groups for data objects using a counter score updated through periodic scan as the metric. The major difference between their strategy and our protocol is that while forming groups they rely on a single score of the peer whereas our protocol disseminates scores across timeline to better capture the cyclic behavior. However, the counter mechanism cannot consistently capture phase relationships within and between peers, e.g., the fact that if a host has diurnal availability, then it will be online for the longest consecutive stretch starting in the morning, when its counter is lowest. Bustamante et al. [24] try to keep friendship with a peer selected among its known ones using peers lifespan distribution. Blond et al. [17], propose two availability-aware applications that takes into account the peers' previous availability pattern collected through an epidemic protocol. Using a simple predictor, they propose to find the best matching peer to meet the specific goal of the application. Sacha et al. [19] try to solve the problem of super-peer [25] selection for naming service with the help of gradient topology using a gossip based method. Their strategy mainly focuses to create a core of super-peers based on a single score of current uptime. A group based chord model is proposed in [27] to minimize the influence of frequent arrivals and departures of peers. This work mainly focuses on routing mechanism after forming groups by simple hashing of IDs.

There have been significant work to improve the performance of replication with a goal to provide better availability. Bolosky et at. [9] utilize machines' uptime, downtime, lifetime and correlation among them them to suggest a replica management algorithm. Douceur et al. investigate a family of randomized, swap-based, hill-climbing algorithms for replica placement from theoretic [22], simulation [10], competitive [21] perspective using an analytic model of machine availability. They show the disadvantages of initial replica placement in an availability-sensitive fashion and suggest an algorithm with random initial placement followed by incremental improvement [20]. Time-related replication [3] uses peers' recent session time to determine the number and placement of replicas optimally. Kavitha et al. [11] proposes a probabilistic method to determine the same. Shi et al. [8] suggest an application specific availability measurement and a two-level DHT to improve that. Mickens et al. [7] proposes three availability prediction mechanisms to improve routing in delay-tolerant networks. Bhagwan et al. [2] explore the issues of replication granularity, replica placement, and application characteristics. In [4] they propose TotalRecall, a system that automatically estimates the availability of hosts, predicts their future availability based on past behavior, calculates the appropriate redundancy mechanisms. They suggest two repair policies: reactive and proactive. In the reactive approach the system reacts to a host going down by replicating its data elsewhere at the cost of additional communication overhead. In the proactive approach of [6],[5], the system continuously monitors the data availability and replicates data in advance when it predicts that the number of replica may fall below the required number. The proactive scheme requires some kind of estimator which works based on some priori knowledge of failure behavior or host availability. Our work falls in the proactive category as it try to ensure persistent storage utilizing previous history of availability.

# 6 Conclusion

In this paper we have introduced a new grouping mechanism that irrespective of peer timing and churn ensures data availability around the clock. Our technique exploits the diurnal pattern of availability exhibited by the globally scattered peers. We successfully verify that our unstructured routing mechanism without any centralized mechanism can accumulate enough information to frame a virtual assemblage of peers as replication points. Each virtual assembly or group together ensures high data availability while keeping the group size, as well as the number of replicas small. To further improve availability while keeping the group size small, we plan to refine our technique and to store availability information using a Distributed Hash Table, which should result into a globally optimized group formation algorithm. Furthermore as future work, we intend to investigate security issues related to group formation and to ensure data availability even in the presence of malicious peers.

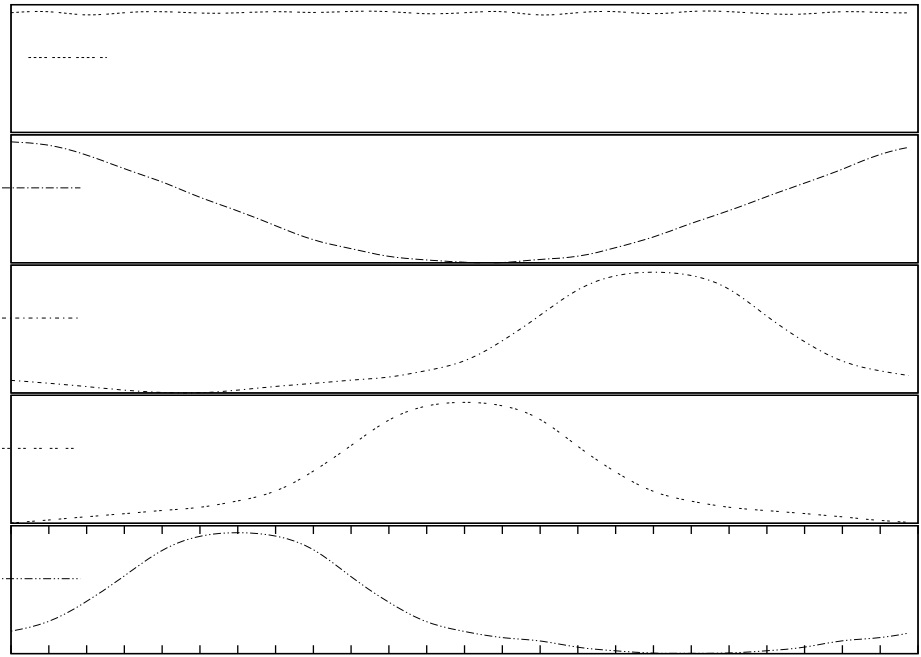